\documentclass[
]{ceurart}

\usepackage{float}
\usepackage{listings}
\usepackage{appendix}
\usepackage{graphicx}
\usepackage{amsmath}
\usepackage{semantic}
\usepackage{hyperref}
\usepackage{xspace}
\usepackage[frozencache=true,cachedir=minted-cache]{minted}

\newtheorem{definition}{Definition}
\newtheorem{assumption}{Assumption}

\newcommand{\Seq}{\textsf{Seq}\xspace}
\newcommand{\NSeq}{\textsf{NSeq}\xspace}

\DeclareMathOperator{\fst}{f}
\DeclareMathOperator{\lst}{l}

\DeclareMathOperator{\set}{set}
\DeclareMathOperator{\get}{get}

\DeclareMathOperator{\const}{const}
\DeclareMathOperator{\relocate}{relocate}
\DeclareMathOperator{\concat}{concat}
\DeclareMathOperator{\slice}{slice}
\DeclareMathOperator{\update}{update}

\DeclareMathOperator{\ite}{ite}
\DeclareMathOperator{\llet}{let}

\begin{document}

\copyrightyear{2024}
\copyrightclause{Copyright for this paper by its authors.
  Use permitted under Creative Commons License Attribution 4.0
  International (CC BY 4.0).}

\conference{SMT 2024: 22nd International Workshop on Satisfiability Modulo Theories}

\title{An SMT Theory for n-Indexed Sequences}
%\todo{error when using textit or emph on the "n". Find a way to put italized n in title}

%\tnotetext[1]{You can use this document as the template for preparing your
%  publication. We recommend using the latest version of the ceurart style.}

\author[1,2]{Hichem Rami {Ait El Hara}}[%
  orcid=0000-0001-7909-0413,
  email=hra687261@gmail.com,
  url=https://hra687261.github.io/,
]

\author[2]{François Bobot}[%
  orcid=0000-0002-6756-0788,
  email=francois.bobot@ocamlpro.com, %url=https://kmitd.github.io/ilaria/,
]

\author[1]{Guillaume Bury}[%
  orcid=0009-0002-1267-251X,
  email=gbury@gmail.com,
  url=https://gbury.eu/,
]

\address[1]{OCamlPro, Paris, France}
\address[2]{Université Paris-Saclay, CEA, List, F-91120, Palaiseau, France}

%\fntext[1]{These authors contributed equally.}

\begin{abstract}
  The SMT (Satisfiability Modulo Theories) theory of arrays is well-established and widely used, with various decision procedures and extensions developed for it. However, recent works suggest that developing tailored reasoning for some theories, such as sequences and strings, is more efficient than reasoning over them through axiomatization over the theory of arrays. In this paper, we are interested in reasoning over n-indexed sequences as they are found in some programming languages, such as Ada. We propose an SMT theory of n-indexed sequences and explore different ways to represent and reason over n-indexed sequences using existing theories, as well as tailored calculi for the theory.
\end{abstract}

% \begin{keywords}
%   Satisfiability Modulo Theories \sep
%   Sequences \sep
%   Arrays
% \end{keywords}

\maketitle
\section{Introduction}

In the SMT theory of sequences, sequences are viewed as a generalization of strings to non-character elements, with possibly infinite alphabets. Sequences are dynamically sized, and their theory has a rich signature. It allows selecting elements of a sequence by their index, concatenating two sequences, extracting sub-sequences, and performing other operations. The expressiveness of the theory of sequences makes it easier to represent various commonly found data structures in programming languages, such as arrays in the C language, lists in Python, etc.

The theory of arrays is less expressive as it only supports selecting and storing one value at one index at a time, and arrays have fixed sizes determined by the number of inhabitants of the sort of indices. In contrast, sequences have dynamic lengths and operations allowing the selection and updating of sets of indices at a time. To use the theory of arrays to represent sequences, one would need to extend it and axiomatize the necessary properties, such as dynamic length and additional operations like concatenation and extraction.

We are interested in a variant of the theory of sequences, which we call the theory of n-indexed sequences. They differ from sequences mainly in their indexing, as they are not necessarily 0-indexed but n-indexed, as their name suggests. This means they are defined as ordered collections of values of the same sort indexed from a first index $n$ to a last index $m$. Such sequences are present in some programming languages like Ada. Since there is no dedicated theory for such sequences, reasoning over them cannot be done straightforwardly using the existing theories of arrays and sequences. It is therefore necessary to use extensions and axiomatizations to reason over them.

In this paper, we will present the theory of n-indexed sequences, its signature and semantics, as well as different ways to reason over it using existing theories and by adapting calculi from the theory of sequences to the theory of n-indexed sequences.

\subsubsection*{Related work:}

The SMT theory of sequences was introduced by Bjorner et al. \cite{bjorner_smt-lib_2012}. Several contributions explored this theory, its syntax and semantics \cite{ait_el_hara_smt_2024}, and its decidability \cite{furia_whats_2010, jez_decision_2023}.

Our theory of n-indexed sequences and the calculi we developed are based on the contribution by Sheng et al. \cite{sheng_reasoning_2023}, which in turn is based on reasoning about the theories of strings \cite{hutchison_dpllt_2014, berzish_z3str3_2017} and arrays \cite{christ_weakly_2015}. Other contributions have extended the theory of arrays with properties that are present in sequences, such as length \cite{bonacina_cdsat_2022, ghilardi_interpolation_2023, bradley_whats_2006} and a concatenation function \cite{wang_solver_2023}.

\section{Notation}

We refer to the theory of n-indexed sequences as the theory of n-sequences or the \NSeq theory, and n-indexed sequence terms will be referred to as n-sequences, n-sequence terms, or \NSeq terms. Their sort will be denoted as \textsf{NSeq E}, where \textsf{E} is the sort of the elements of the n-sequence. We will refer to the theory of sequences as the \Seq theory. \textsf{Int} is the sort of integers from the theory of Linear Integer Arithmetic. $\min$ and $\max$ are the usual mathematical functions. $\ite$ is the $\ite$ SMT-LIB function; it takes a boolean expression and two expressions of the same sort and returns the first one if the boolean expression is true and the second otherwise. The $\llet\ x = v, y$ symbol binds a variable $x$ to a value $v$ in a term $y$.

In the remainder of the paper, we will use $s$, $s_n$, $w$, $w_n$, $y_1$, $y_2$, $z_1$, and $z_2$ to represent \NSeq terms, with $n$ being a positive integer. $k$, $k_1$, $k_2$, and $k_3$ will represent fresh \NSeq term variables. $i$ and $j$ will be used as integer index terms, and $u$ and $v$ will be used as \NSeq element terms. We assume that all the terms we use are well-sorted.

\section{The Theory of n-Indexed Sequences}

\begin{table}[ht]
  \caption{The signature of the theory of n-indexed sequences}
  \label{fig:nseq_signature}
  \centering
  \begin{tabular}{p{0.3\linewidth} p{0.6\linewidth}}
    \hline
    SMT-LIB symbol         & Sort                                                                           \\
    \hline
    \textsf{nseq.get}      & \textsf{NSeq E \(\rightarrow\) Int \(\rightarrow\) E}                          \\
    \textsf{nseq.set}      & \textsf{NSeq E \(\rightarrow\) Int \(\rightarrow\) E \(\rightarrow\)  NSeq E}  \\
    \textsf{nseq.first}    & \textsf{NSeq E \(\rightarrow\) Int}                                            \\
    \textsf{nseq.last}     & \textsf{NSeq E \(\rightarrow\) Int}                                            \\
    \textsf{nseq.const}    & \textsf{Int \(\rightarrow\) Int \(\rightarrow\) E \(\rightarrow\)  NSeq E}     \\
    \textsf{nseq.relocate} & \textsf{NSeq E \(\rightarrow\) Int \(\rightarrow\) NSeq E}                     \\
    \textsf{nseq.concat}   & \textsf{NSeq E \(\rightarrow\) NSeq E \(\rightarrow\) NSeq E}                  \\
    \textsf{nseq.slice}    & \textsf{NSeq E \(\rightarrow\) Int \(\rightarrow\) Int \(\rightarrow\) NSeq E} \\
    \textsf{nseq.update}   & \textsf{NSeq E \(\rightarrow\) NSeq E \(\rightarrow\) NSeq E}                  \\
    \hline
  \end{tabular}
\end{table}

We present in this section the theory of n-indexed sequences. The signature of the \NSeq theory is presented in the table \ref{fig:nseq_signature}. In the remainder of the paper, when referring to the symbols of the theory, the prefix "\textsf{nseq.}" of the symbols will be omitted.

The following list describes the semantics of each symbol of the theory:

\begin{itemize}
  \item $\fst_s$: the first index of $s$.
  \item $\lst_s$: the last index of $s$.
  \item $\get(s, i)$: if the index $i$ is within the bounds of $s$, returns the element associated with $i$ in $s$; otherwise, returns an uninterpreted value. An uninterpreted value is one which is not constrained and can be any value of the right sort.
  \item $\set(s, i, v)$: if $i$ is within the bounds of $s$, creates a new n-sequence in which $i$ is associated with $v$, and the other indices within the bounds are associated with the same values as the corresponding indices in $s$; otherwise, returns $s$.
  \item $\const(f, l, v)$: creates an n-sequence with $f$ as the first index and $l$ as the last, and if it is not empty, all the indices within its bounds are associated with the value $v$.
  \item $\relocate(a, f)$: given an n-sequence $a$ and an index $f$, returns a new n-sequence $b$ which has $f$ as its first index and $(f + \lst_a - \fst_a)$ as its last index, and associates with each index $i$ within the bounds of $b$ the same value associated with the index $(i - \fst_b + \fst_a)$ in $a$.
  \item $\concat(a, b)$: if $a$ is empty, returns $b$; if $b$ is empty, returns $a$; if $\fst_b = \lst_a + 1$, returns a new n-sequence in which the first index is $\fst_a$, the last index is $\lst_b$, and all the indices within the bounds of $a$ are associated with the same values as the corresponding indices in $a$, as well as the indices within the bounds of $b$ are associated with the same values as the corresponding indices in $b$; if $\fst_b \neq \lst_a + 1$, returns $a$.
  \item $\slice(a, f, l)$: if $\fst_a \le f \le l \le \lst_a$, returns a new n-sequence for which the first index is $f$, the last index is $l$, and all the indices between $f$ and $l$ are associated with the same values as the corresponding indices in $a$; otherwise, returns $a$.
  \item $\update(a, b)$: if $a$ is empty, or $b$ is empty, or the property $\fst_a \le \fst_b \le \lst_b \le \lst_a$ doesn't hold, returns $a$; otherwise, returns a new n-sequence $c$ such that $\fst_c = \fst_a$ and $\lst_c = \lst_a$, and for all the indices within the bounds of $c$, if they are within the bounds of $b$, then they are associated with the same values as they are in $b$, otherwise they are associated with the values to which they are associated in $a$.
\end{itemize}

\begin{definition}[Bounds]
  The bounds of an n-sequence $s$ are its first and last index, which are respectively denoted as $\fst_s$ and $\lst_s$, and which correspond to the values returned by the functions $\textsf{nseq.first}(s)$ and $\textsf{nseq.last}(s)$ respectively. An index $i$ is said to be within the bounds of an n-sequence $s$ if:
  $$ \fst_s \le i \le \lst_s $$
\end{definition}

\begin{definition}[Extensionality]
  The theory of n-indexed sequences is extensional, which means that n-sequences that contain the same elements are equal. Therefore, given two n-sequences $a$ and $b$:

  \[
    \begin{array}{c}
      \fst_a = \fst_b \land \lst_a = \lst_b \land                                            \\
      (\forall i: \textsf{Int}, \fst_a \le i \le \lst_a \rightarrow \get(a, i) = \get(b, i)) \\
      \rightarrow a = b
    \end{array}
  \]
\end{definition}

\begin{definition}[Empty n-sequence]
  An n-sequence $s$ is said to be empty if $\lst_s < \fst_s$. Two empty n-sequences $a$ and $b$ are equal if $\fst_a = \fst_b$ and $\lst_a = \lst_b$; otherwise, they are distinct.
\end{definition}

\section{Reasoning with existing theories}

One way to reason over the \NSeq theory is by using the theory of arrays. It is done by extending it with the symbols of the \NSeq theory and adding the right axioms that follow the semantics of the corresponding symbols in the \NSeq theory.

Another way is to use the theory of sequences and the theory of algebraic data types. It consists in defining n-sequences as a pair of a sequence and the first index (the offset to zero):

\begin{minted}[frame=single]{scheme}
(declare-datatype NSeq
  (par (T) ((nseq.mk (nseq.first Int) (nseq.seq (Seq T))))))
\end{minted}

The other symbols of the \NSeq theory can also be defined using the \NSeq data type defined above, for example:

\begin{minted}[frame=single]{scheme}
(define-fun nseq.last (par (T) ((s (NSeq T))) Int
  (+ (- (seq.len (nseq.seq s)) 1) (nseq.first s))))

(define-fun nseq.get (par (T) ((s (NSeq T)) (i Int)) T
  (seq.nth (nseq.seq s) (- i (nseq.first s)))))

(define-fun nseq.set (par (T) ((s (NSeq T)) (i Int) (v T)) (NSeq T)
  (nseq.mk (nseq.first s)
    (seq.update (nseq.seq s) (- i (nseq.first s)) (seq.unit v)))))
\end{minted}

Except for the $\const$ function which needs to be axiomatized:

\begin{minted}[frame=single]{scheme}
(declare-fun nseq.const (par (T) (Int Int T) (NSeq T)))

;; "nseq_const"
(assert (par (T) (forall ((f Int) (l Int) (v T))
    (!
      (let ((s (nseq.const f l v)))
        (and
          (= (nseq.first s) f)
          (= (nseq.last s) l)
          (forall ((i Int))
            (=> (and (<= f i) (<= i l)) (= (nseq.get s i) v)))))
      :pattern ((nseq.const f l v))))))
\end{minted}

The full \NSeq theory, defined using the \Seq theory and Algebraic Data Types, is attached in Appendix \ref{appendix:seq_adt}.

Although this approach allows us to reason over n-indexed sequences, it is not ideal to depend on two theories to do so. Additionally, the differences in semantics between the $\update$ and $slice$ functions of the \NSeq theory, and the $\textsf{seq.update}$ and $\textsf{seq.extract}$ functions of the \Seq theory, make the definitions relatively complex.

\section{Porting Calculi from the \Seq Theory to the \NSeq Theory}
\label{section:calculi}

To develop our calculi over the \NSeq theory, we based our work on the calculi developed by Sheng et al. \cite{sheng_reasoning_2023} on the \Seq theory, where two calculi were proposed. The first is called the \textsf{BASE} calculus, based on a string theory calculus that reduces functions selecting and storing one element at an index to concatenations of sequences. The second is called the \textsf{EXT} calculus, and it handles these functions using array-like reasoning. Our versions of these calculi are referred to as \textsf{NS-BASE} and \textsf{NS-EXT}, respectively.

The \NSeq theory differs from the \Seq theory in both syntax and semantics of many symbols:

\begin{itemize}
  \item $\const$ and $\relocate$ do not appear in the \Seq theory, while \textsf{seq.empty}, \textsf{seq.unit}, and \textsf{seq.len} do not appear in the \NSeq theory.
  \item The \textsf{seq.nth} function corresponds to the $\get$ function in the \NSeq theory.
  \item \textsf{seq.update} from the \Seq theory with a value as a third argument corresponds to $\set$ in the \NSeq theory, while \textsf{seq.update} with a sequence as a third argument corresponds to $\update$ in the \NSeq theory, which takes only two \NSeq terms as arguments.
  \item \textsf{seq.extract} in the \Seq theory takes a sequence, an offset, and a length, and corresponds to $\slice$ in the \NSeq theory, which takes an n-sequence, a first index, and a last index.
  \item The concatenation function (\textsf{seq.++}) in \Seq is n-ary, and it corresponds to $\concat$ in the \textsf{NSeq} theory, which is binary.
\end{itemize}

Therefore, we needed to make substantial changes to the \Seq theory calculi to adapt them to the \NSeq theory. In this section, we present the resulting calculi. We assume that we are in a theory combination framework where reasoning over the LIA (Linear Integer Arithmetic) theory is supported, and where unsatisfiability in one of the theories implies unsatisfiability of the entire reasoning. We will only present the rules that handle the symbols of the \NSeq theory.

\subsection{Common calculus}

\begin{definition}[Equivalence modulo relocation]
  Given two \NSeq terms $s_1$ and $s_2$, the terms are said to be equivalent modulo relocation, denoted with the equivalence relation $s_1 =_{reloc} s_2$, such that:
  \[
    \begin{array}{c}
      s_1 =_{reloc} s_2 \equiv \\
      \lst_{s_2} = \lst_{s_1} - \fst_{s_1} + \fst_{s_2} \land \forall i: Int, \fst_{s_1} \le i \le \lst_{s_1} \Rightarrow get(s_1, i) = get(s_2, i - \fst_{s_1} + \fst_{s_2} )
    \end{array}
  \]
  Equivalence modulo relocation represents equivalence between n-sequences relative to their starting index. Two n-sequences are equivalent modulo relocation if they have the same set of elements in the same order, but start at different indices.
\end{definition}

\begin{definition}[\NSeq term normal form]
  For simplicity and consistency with \Seq theory calculi, we introduce the concatenation operator $::$ with the invariant:
  \[
    \begin{array}{c c c}
      s = s_1 :: s_2 & \implies & \fst_s = \fst_{s_1} \land \lst_s = \lst_{s_2} \land \fst_{s_2} = \lst_{s_1} + 1
    \end{array}
  \]
  This operator is used to normalize \NSeq terms. It differs from $\concat$ by not having to check the condition that $\fst_{s_2} = \lst_{s_1} + 1$ before concatenation, as it is ensured by the invariant.
\end{definition}

\begin{assumption}
  \label{assumption:rewrites}
  We assume that the following simplification rewrites are applied whenever possible:
  \[
    \begin{array}{c c c l l}
      s_1 :: s_2 & \rightarrow & s_1                      & \text{when } \lst_{s_2} < \fst_{s_2} & (1) \\
      s_1 :: s_2 & \rightarrow & s_2                      & \text{when } \lst_{s_1} < \fst_{s_1} & (2) \\
      s_1 :: s_2 & \rightarrow & s_1 :: w_1 :: ... :: w_n & \text{when } s_2 = w_1 :: ... :: w_n & (3) \\
      s_1 :: s_2 & \rightarrow & w_1 :: ... :: w_n :: s_2 & \text{when } s_1 = w_1 :: ... :: w_n & (4)
    \end{array}
  \]
  $(1)$ and $(2)$ remove empty \NSeq terms from normal form. $(3)$ and $(4)$ make sure that when an \NSeq term appear in the normal of another \NSeq term and has its own normal form, then it is replaced by its normal form.
\end{assumption}

\begin{figure}[!htb]
  \centering
  \[
    \begin{array}{c}
      \inference[\textsf{Const-Bounds}] {s = \const(f,l,v)} {
        \begin{array}{c c}
          \fst_s = f \land \lst_s = l
        \end{array}}
      \quad
      \inference[\textsf{Reloc-Bounds}] {s_1 = relocate(s_2, i)} {
        \begin{array}{c c}
          i = \fst_{s_2} \land s_1 = s_2 & ||                                                   \\
          i \neq \fst_{s_2} \land \fst_{s_1} = i \land \lst_{s_1} = i + \lst_{s_2} - \fst_{s_2} \\
          \land s_1 =_{reloc} s_2
        \end{array}}
      \\\\

      \inference[\textsf{NS-Slice}] {s_1 = slice(s, f, l)} {
        \begin{array}{c c}
          (f < \fst_{s} \lor l < f \lor \lst_{s} < l) \land s_1 = s & || \\
          \fst_{s} \le f \le l \le \lst_{s} \land s = k_1 :: s_1 :: k_2
        \end{array}}
      \\\\

      \inference[\textsf{NS-Concat}] {s = concat(s_1, s_2)} {
        \begin{array}{c c}
          \lst_{s_1} < \fst_{s_1} \land s = s_2                                       & || \\
          (\lst_{s_2} < \fst_{s_2} \lor \lst_{s_1} + 1 \neq \fst_{s_2}) \land s = s_1 & || \\
          \fst_{s_1} \le \lst_{s_1} \land \fst_{s_2} \le \lst_{s_2} \land
          \fst_{s_2} = \lst_{s_1} + 1 \land s = s_1 :: s_2
        \end{array}}
      \\\\

      \inference[\textsf{NS-Update}] {s_1 = update(s_2, s)} {
        \begin{array}{c c}
          (\lst_{s} < \fst_{s} \lor f_s < \fst_{s_2} \lor \lst_{s_2} < \lst_{s}) \land s_1 = s_2 & || \\
          \fst_{s_2} \le \fst_{s} \le \lst_{s} \le \lst_{s_2} \land s_1 = k_1 :: s :: k_3 \land s_2 = k_1 :: k_2 :: k_3
        \end{array}}
      \\\\

      \inference[\textsf{NS-Split}] {s = w :: y_1 :: z_1                 & s = w :: y_2 :: z_2} {
        \begin{array}{c l}
          \lst_{y_1} = \lst_{y_2} \land y_1 = y_2
           & || \\
          \lst_{y_1} > \lst_{y_2} \land y_1 = y_2 :: k \land \fst_{k} = \lst_{y_2} + 1 \land \lst_{k} = \lst_{y_1}
           & || \\
          \lst_{y_1} < \lst_{y_2} \land y_2 = y_1 :: k \land \fst_{k} = \lst_{y_1} + 1 \land \lst_{k} = \lst_{y_2}
        \end{array}}
      \\\\

      \inference[\textsf{NS-Comp-Reloc}] {s_1 = k_1 :: k_2 :: ... :: k_n & s_1 =_{reloc} s_2} {
      \begin{array}{c l}
        \fst_{s_1} = \fst_{s_2} \land s_1 = s_2 & ||                                                  \\
        s_2 = relocate(k_1, \fst_{s_2}) :: relocate(k_2, \fst_{k_2} - \fst_{s_1} + \fst_{s_2}) :: ... \\ :: relocate(k_n, \fst_{k_n} - \fst_{s_1} + \fst_{s_2})
      \end{array}}
    \end{array}
  \]
  \caption{Common inference rules for the \textsf{NS-BASE} and \textsf{NS-EXT} calculi}
  \label{fig:common_rules}
\end{figure}

Figure \ref{fig:common_rules} illustrates a set of common rules shared between the two calculi \textsf{NS-BASE} and \textsf{NS-EXT}. The rules \textsf{Const-Bounds} and \textsf{Reloc-Bounds} propagate the bounds of constant and relocated n-sequences, which are created using the $\const$ and $\relocate$ functions, respectively. The rules \textsf{NS-Slice}, \textsf{NS-Concat}, and \textsf{NS-Update} handle $\slice$, $\concat$, and $\update$ by normalizing the \NSeq terms under appropriate conditions. If an \NSeq term has two normal forms where distinct terms begin at the same index but end at different indices, the \textsf{NS-Split} rule rewrites the longer term as a concatenation of the shorter one and a fresh variable. The \textsf{NS-Comp-Reloc} rule propagates concatenations over equivalence modulo relocation.

\subsection{The base calculus}

\begin{figure}[!htb]
  \centering
  \[
    \begin{array}{c}
      \inference[\textsf{R-Get}] {v = get(s, i)} {
        \begin{array}{c c}
          i < \fst_{s} \lor \lst_{s} < i \quad & || \\
          \fst_{s} \le i \le \lst_{s} \land s = k_1 :: const(i, i, v) :: k_2
        \end{array}}
      \\\\

      \inference[\textsf{R-Set}] {s_1 = set(s_2, i, v)} {
        \begin{array}{c c}
          (i < \fst_{s_2} \lor \lst_{s_2} < i) \land s_1 = s_2
           & ||                                                                                       \\
          \fst_{s_2} \le i \le \lst_{s_2} \land \fst_{s_1} = \fst_{s_2} \land \lst_{s_1} = \lst_{s_2} \\
          s_1 = k_1 :: const(i, i, v) :: k_3 \land s_2 = k_1 :: k_2 :: k_3
        \end{array}
      }
    \end{array}
  \]
  \caption{\textsf{NS-BASE} specific inference rules}
  \label{fig:nsbase_rules}
\end{figure}

The base calculus comprises the rules in figures \ref{fig:common_rules} and \ref{fig:nsbase_rules}. The rules \textsf{R-Get} and \textsf{R-Set} handle the $get$ and $set$ operations by introducing a new normal form for the \textsf{NSeq} terms they operate on. In the \textsf{R-Get} rule, when $i$ is within the bounds of $s$, a new normal form of $s$ is introduced. It includes a constant \textsf{NSeq} term of size one at the $i$th position storing the value $v$, and two variables, $k_1$ and $k_2$, to represent the left and right segments of \textsf{NSeq} term $s$ respectively. The \textsf{R-Set} rule operates similarly: when $i$ is within the bounds of $s_2$, new normal forms are introduced for both $s_1$ and $s_2$. These forms share two variables, $k_1$ and $k_3$, representing segments on the left and right of the $i$th index. $s_1$ has a constant \textsf{NSeq} term of size one holding the value $v$ at the $i$th index, while $s_2$ introduces another variable, $k_2$, of the same size and position.

\subsection{The extended calculus}

\begin{figure}[!htb]
  \centering
  \[
    \begin{array}{c}
      \inference[\textsf{Get-Concat}] {v = get(s, i)            & s = w_1 :: ... :: w_n} {
        \begin{array}{c l}
          i < \fst_{s} \lor \lst_{s} < i                        & ||                    \\
          \fst_{w_1} \le i \le \lst_{w_1} \land get(w_1, i) = v & || \quad ... \quad || \\
          \fst_{w_n} \le i \le \lst_{w_n} \land get(w_n, i) = v
        \end{array}}
      \\\\

      \inference[\textsf{Set-Concat}] {s_1 = set(s_2, i, v)     & s_2 = w_1 :: ... :: w_n} {
        \begin{array}{c l}
          i < \fst_{s_2} \lor \lst_{s_2} < i
                                                                                                                                                         & ||
          \\
          s_1 = k_1 :: ... :: k_n \land \fst_{w_1} \le i \le \lst_{w_1} \land k_1 = set(w_1, i, v)             \land
                                                                                                                                                         &
          \\
          \fst_{k_1} = \fst_{w_1} \land \lst_{k_1} = \lst_{w_1} \land ... \land                    \fst_{k_n} = \fst_{w_n} \land \lst_{k_n} = \lst_{w_n} & || \quad ... \quad ||
          \\
          s_1 = k_1 :: ... :: k_n \land \fst_{w_n} \le i \le \lst_{w_n} \land k_n = set(w_n, i, v) \land                                                 &
          \\
          \fst_{k_1} = \fst_{w_1} \land \lst_{k_1} = \lst_{w_1} \land ... \land                    \fst_{k_n} = \fst_{w_n} \land \lst_{k_n} = \lst_{w_n}
        \end{array}}
      \\\\

      \inference[\textsf{Set-Concat-Inv}] {s_1 = set(s_2, i, v) & s_1 = w_1 :: ... :: w_n} {
        \begin{array}{c l}
          i < \fst_{s_2} \lor \lst_{s_2} < i
           & ||                 \\
          s_2 = k_1 :: ... :: k_n \land \fst_{w_1} \le i \le \lst_{w_1} \land w_1 = set(k_1, i, v) \land
           &                    \\
          \fst_{k_1} = \fst_{w_1} \land \lst_{k_1} = \lst_{w_1} \land ... \land                    \fst_{k_n} = \fst_{w_n} \land \lst_{k_n} = \lst_{w_n}
           &
          || \quad ... \quad || \\
          s_2 = k_1 :: ... :: k_n \land \fst_{w_n} \le i \le \lst_{w_n} \land w_n = set(k_n, i, v) \land
           &                    \\
          \fst_{k_1} = \fst_{w_1} \land \lst_{k_1} = \lst_{w_1} \land ... \land                    \fst_{k_n} = \fst_{w_n} \land \lst_{k_n} = \lst_{w_n}
        \end{array}}
      \\\\

      \inference[\textsf{Get-Const}] {s = const(f, l, v)        & u = get(s, i)} {
        \begin{array}{c}
          i < \fst_{s} \lor \lst_{s} < i \quad || \quad \fst_{s} \le i \le \lst_{s} \land u = v
        \end{array}}
      \\\\

      \inference[\textsf{Get-Intro}] {s_1 = set(s_2, i, v)} {
        \begin{array}{c}
          i < \fst_{s_1} \lor \lst_{s_1} < i \quad || \quad \fst_{s_1} \le i \le \lst_{s_1} \land v = get(s_1, i) \quad
        \end{array}}
      \\\\

      \inference[\textsf{Get-Set}] {s_1 = set(s_2, i, v)        & u = get(s_1, j)} {
        \begin{array}{c l}
          i < \fst_{s_1} \lor \lst_{s_1} < i                      & || \\
          i = j \land \fst_{s_1} \le i \le \lst_{s_1} \land u = v & || \\
          i \neq j \land \fst_{s_1} \le i \le \lst_{s_1} \land u = get(s_2, i)
        \end{array}}
      \\\\

      \inference[\textsf{Set-Bound}] {s_1 = set(s_2, i, v)} {
        \begin{array}{c}
          s_1 = s_2 \quad || \quad \fst_{s_1} \le i \le \lst_{s_1} \land get(s_2, i) \neq v
        \end{array}}
      \\\\

      \inference[\textsf{Get-Reloc}] {v = get(s_1, i)           & s_1 =_{reloc} s_2} {
      \begin{array}{c}
        i < f_s \lor l_s < i \quad || \quad
        f_s \le i \le l_s \land v = get(s_2, i - \fst_{s_1} - \fst_{s_2})
      \end{array}}
    \end{array}
  \]
  \caption{\textsf{NS-EXT} specific inference rules}
  \label{fig:nsext_rules}
\end{figure}

The extended calculus consists of the rules in figures \ref{fig:common_rules} and \ref{fig:nsext_rules}. It differs from the base calculus by handling the $\get$ and $\set$ functions similarly to how they are treated in the array decision procedure described in \cite{christ_weakly_2015}. The \textsf{Get-Intro} rule introduces a $\get$ operation from a $\set$ operation. The \textsf{Get-Set} operation is equivalent to what is commonly referred to as the \textsf{read-over-write} or \textsf{select-over-store} rule in the \textsf{Array} theory, allowing the application of a $\get$ operation over a $\set$ operation. The \textsf{Set-Bound} rule ensures that a $\set$ operation is performed within the bounds of the target \textsf{NSeq} term, or that the resulting \textsf{NSeq} term is equivalent to the one it was applied on. The \textsf{Get-Concat}, \textsf{Set-Concat}, and \textsf{Set-Concat-Inv} rules illustrate how $\get$ and $\set$ operations are handled when applied to an \textsf{NSeq} term in normal form, where the operations affect the right component of the concatenation within its bounds. The \textsf{Get-Const} rule addresses the special case where a $\get$ operation is applied to a constant \NSeq term. The \textsf{Get-Reloc} rule facilitates the propagation of constraints on index-associated values in \textsf{NSeq} terms from one term to others that are equivalent modulo relocation.

\section{Implementation}

We have implemented a prototype of the described calculi in the Colibri2 CP (Constraint Programming) solver. In this section, we discuss some of the implementation choices we made.

The rewriting rules described in Assumption \ref{assumption:rewrites} are applied whenever applicable using a callback system. When the conditions are satisfied, the corresponding rewriting rule is triggered.

Equivalence modulo relocation is managed using a disjoint-set (union-find) data structure. In this  data structure, the elements of the sets are \textsf{NSeq} terms, and the equivalence relation is defined by $=_{reloc}$ as previously specified. By definition, if two elements of an equivalence class are at the same relocation offset from the representative, they are equal. The data structure maintains, for each equivalence class, a mapping from offset to an element of the class that is at that specific relocation offset from the representative. This facilitates efficient detection of such equalities with minimal overhead.

\section{Experimental Results}

In this section, we present experimental results of the calculi described in the previous section. Currently, our experiments have focused exclusively on quantifier-free benchmarks that utilize only the theory of sequences and the theory of uninterpreted functions. These benchmarks constitute a subset of those employed in the paper \cite{sheng_reasoning_2023}, originally translated into the \Seq theory from the \textsf{QF\_AX} \textsf{SMT-LIB} benchmarks.

To achieve this, we implemented support for the \Seq theory in our solver by translating \Seq terms into \NSeq terms. The translation process is as follows:

\begin{itemize}
  \item \Seq terms: \NSeq terms for which the first index is 0 and the last index is greater or equal than $-1$.
  \item \textsf{seq.empty}: an \NSeq term of the same sort, in which the first index is 0 and the last is $-1$, denoted $\epsilon$.
  \item $\textsf{seq.unit}(v)$: $\const(0,0,v)$
  \item $\textsf{seq.len}(s)$: $\lst_s - \fst_s + 1$
  \item $\textsf{seq.nth}(s, i)$: $\get(s, i)$
  \item $\textsf{seq.update}(s_1, i, s_2)$:
        \[
          \begin{array}{c}
            \llet (r, \relocate(s_2, i),
            \ite(\fst_{s_1} \le i \le \lst_{s_1} \land \lst_{s_1} < \lst_r, \\
              \update(s_1, \slice(r, i, \lst_{s_1})),
              \update(s_1, r)))
          \end{array}
        \]
  \item $\textsf{seq.extract}(s, i, j)$:
        \[
          \begin{array}{c}
            \ite(i < \fst_s \lor \lst_s < i \lor j \le 0, \epsilon, \slice(s, i, \min(\lst_s,i + j - 1)))
          \end{array}
        \]
  \item $\textsf{seq.++}(s_1, s_2, s_3, ..., s_n)$:
        \[
          \begin{array}{c}
            \llet(c_1, \concat(s_1, \relocate(s_2,\lst_{s_1}+1)), \\
            \llet(c_2, \concat(c_1, \relocate(s_3,\lst_{c_1}+1)), \\
            ...                                                   \\
              \concat(c_{n - 2}, \relocate(s_n,\lst_{c_{n - 2}}+1))))
          \end{array}
        \]
\end{itemize}

\begin{figure}
  \centering
  \begin{minipage}{.49\textwidth}
    \centering
    \includegraphics[width=\textwidth]{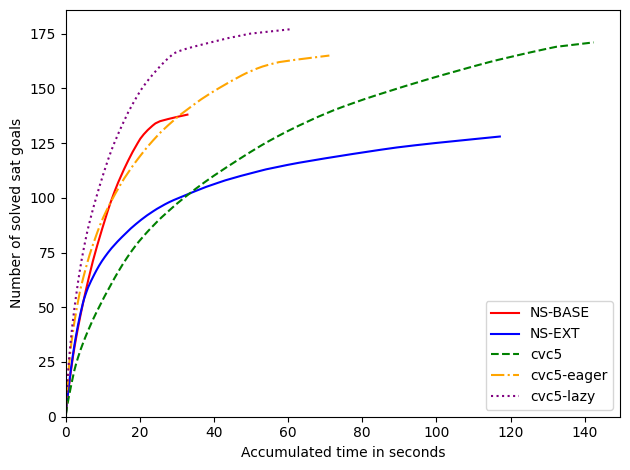}
  \end{minipage}
  \begin{minipage}{.49\textwidth}
    \centering
    \includegraphics[width=\textwidth]{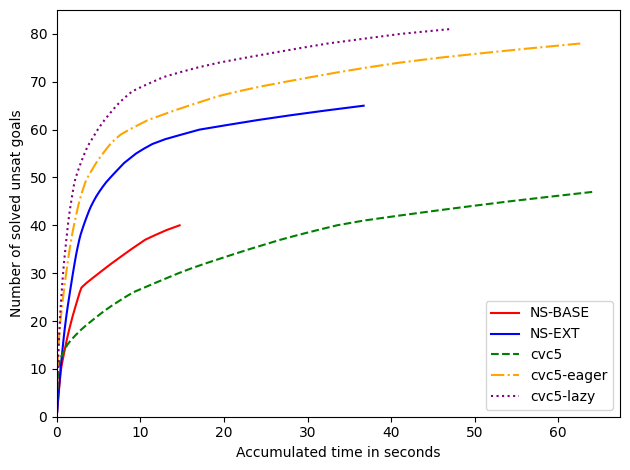}
  \end{minipage}
  \caption{Number of solved goals by accumulated time in seconds on quantifier-free Seq benchmarks translated from the \textsf{QF\_AX} \textsf{SMT-LIB} benchmarks}
  \label{fig:results}
\end{figure}

The figure \ref{fig:results} depicts the number of satisfiable and unsatisfiable goals solved over accumulated time using our prototype implementation and the cvc5 SMT solver with different command-line options. \textsf{NS-BASE} and \textsf{NS-EXT} refer to our implementations\footnote{Available at: \href{https://git.frama-c.com/pub/colibrics/-/tree/smt2024}{https://git.frama-c.com/pub/colibrics/-/tree/smt2024} (commit SHA: 43024e674ef26673d2495f3b186954fa37bc3890)}, described in section \ref{section:calculi}, which can be used by running Colibri2 with the command-line options \verb|--nseq-base| and \verb|--nseq-ext|, respectively. We compare our implementation with the \Seq theory implementation in cvc5 (version 1.1.1). In the graphs in figure \ref{fig:results}, \textsf{cvc5} corresponds to running cvc5 with the command-line option \verb|--strings-exp|, necessary for using cvc5's solver for the \Seq theory. \textsf{cvc5-eager} uses the same option with \verb|--seq-arrays=eager|, and \textsf{cvc5-lazy} with \verb|--seq-arrays=lazy|, these options indicate different strategies for using an array-inspired solver for the \Seq theory.

Examining the graph on the right, which shows performance on unsatisfiable goals, we observe that our \textsf{NS-EXT} implementation outperforms \textsf{cvc5} in both time and number of goals solved. Meanwhile, \textsf{NS-BASE} initially solves more goals than \textsf{cvc5}, but solves fewer overall. Additionally, \textsf{cvc5-eager} and \textsf{cvc5-lazy} solve more goals in less time compared to the others. The same trends apply to the satisfiable case, with the exception that \textsf{cvc5} also surpasses \textsf{NS-EXT} in both time and number of goals solved once the 20-second threshold is reached.

In our context, focused on program verification, the performance on unsatisfiable goals holds greater significance, though the satisfiable case remains useful. Since Colibri2 constructs concrete counterexamples before concluding satisfiability, we aim to enhance our current model generation technique for n-sequences. In the unsatisfiable case, while we compete closely with the state-of-the-art SMT solver cvc5, we have observed that some goals unsolved within a short timeout (5 seconds) also remain unsolved with longer timeouts, suggesting potential performance limitations in our propagators for the \NSeq theory. It's also notable that our translation from \Seq to \NSeq in Colibri2 introduces more complex terms, and Colibri2 lacks clause learning, making decisions costlier than in other SMT solvers.

\section{Conclusion}

In this paper, we explored the topic of n-indexed sequences in SMT. We proposed a theory for such sequences and discussed approaches for reasoning over it, whether by using existing theories or by adapting calculi from the theory of sequences to this theory.

Looking ahead, our future work will delve deeper into different reasoning approaches for this theory, exploring their respective strengths and weaknesses through benchmarking with n-indexed sequences. We aim to prove the correctness of our developed calculi and explore alternative methods for reasoning over n-sequences beyond traditional sequence or string reasoning. Moreover, we seek to identify additional applications for this theory beyond programming languages where n-indexed sequences are present.

\bibliography{theories_doc,smt_lib_bib}

\appendix

\section{Appendix}

\subsection{Representation of n-Indexed Sequences using Sequences and Algebraic Data Types}
\label{appendix:seq_adt}

\begin{minted}[frame=single]{scheme}
(declare-datatypes ((NSeq 1))
  ((par (T) ((nseq.mk (nseq.first Int) (nseq.seq (Seq T)))))))

(define-fun nseq.last (par (T) ((s (NSeq T))) Int
  (+ (- (seq.len (nseq.seq s)) 1) (nseq.first s))))

(define-fun nseq.get (par (T) ((s (NSeq T)) (i Int)) T
  (seq.nth (nseq.seq s) (- i (nseq.first s)))))

(define-fun nseq.set (par (T) ((s (NSeq T)) (i Int) (v T)) (NSeq T)
  (nseq.mk (nseq.first s)
    (seq.update (nseq.seq s) (- i (nseq.first s)) (seq.unit v)))))

(declare-fun nseq.const (par (T) (Int Int T) (NSeq T)))

;; "nseq_const"
(assert (par (T) (forall ((f Int) (l Int) (v T))
    (!
      (let ((s (nseq.const f l v)))
        (and
          (= (nseq.first s) f)
          (= (nseq.last s) l)
          (forall ((i Int))
            (=> (and (<= f i) (<= i l)) (= (nseq.get s i) v)))))
      :pattern ((nseq.const f l v))))))

(define-fun nseq.relocate (par (T) ((s (NSeq T)) (f Int)) (NSeq T)
     (nseq.mk f (nseq.seq s))))

(define-fun nseq.concat (par (T) ((s1 (NSeq T)) (s2 (NSeq T))) (NSeq T)
  (ite (< (nseq.last s1) (nseq.first s1))
    s2
    (ite
      (or
        (< (nseq.last s2) (nseq.first s2))
        (not (= (nseq.first s2) (+ (nseq.last s1) 1))))
      s1
      (nseq.mk
        (nseq.first s1)
        (seq.++ (nseq.seq s1) (nseq.seq s2)))))))

(define-fun nseq.slice (par (T) ((s (NSeq T)) (f Int) (l Int)) (NSeq T)
  (ite
    (and
      (<= f l)
      (and (<= (nseq.first s) f) (<= l (nseq.last s))))
    (nseq.mk f (seq.extract (nseq.seq s) (- f (nseq.first s)) (+ (- l f) 1)))
    s)))

(define-fun nseq.update (par (T) ((s1 (NSeq T)) (s2 (NSeq T))) (NSeq T)
  (ite
    (and
      (<= (nseq.first s2) (nseq.last s2))
      (<= (nseq.first s1) (nseq.first s2))
      (<= (nseq.last s2) (nseq.last s1)))
    (nseq.mk (nseq.first s1)
      (seq.update
        (nseq.seq s1)
        (- (nseq.first s2) (nseq.first s1))
        (nseq.seq s2)))
     s1)))
\end{minted}

\end{document}